\documentclass[%
 aip,
 amsmath,amssymb,
 preprint,
 floatfix,
]{revtex4-1}

\usepackage[dvipsnames]{xcolor}
\usepackage{graphicx}
\usepackage{bm}
\usepackage{tikz}
\usetikzlibrary{arrows.meta,decorations.pathmorphing}

\usepackage[T1]{fontenc}
\usepackage{mathptmx}
\usepackage{etoolbox}

\begin{document}

\title{Integrated 3D fully kinetic simulation of field-reversed-configuration formation with embedded coils}

\author{Bowen Zhu}
\author{Jian Wu}
 \email[Author to whom correspondence should be addressed: ]{jxjawj@mail.xjtu.edu.cn}
\affiliation{State Key Laboratory of Electrical Insulation and Power Equipment,
Xi'an Jiaotong University, Xi'an 710049, China}

\date{\today}

\begin{abstract}
We present an integrated, three-dimensional, fully kinetic particle-in-cell simulation of field-reversed-configuration (FRC) formation at the device scale. To our knowledge, this is the first fully kinetic model of whole-device FRC formation. The model embeds the drive coils directly inside the computational domain as physical conductors, advancing them self-consistently with the plasma on a single explicit grid and coupling them in closed loop to an external circuit. We apply this unified framework to the Yingguang-1 $\theta$-pinch. Unlike the magnetohydrodynamic and hybrid models used previously, our framework advances the electrons as kinetic particles rather than a fluid, capturing fast magnetic reconnection and electron heating from first principles. The simulation reproduces the complete formation sequence, from reversed-bias lock-in through reconnection to the emergence of a closed-flux FRC, reaching a peak ion density ${\sim}2.2\times10^{22}\,\mathrm{m^{-3}}$ consistent with experiment. The compressed core is electron-dominated, with $T_e\approx1.7\,$keV exceeding $T_i\approx1.2\,$keV, and is pinched to a separatrix radius $r_s\approx1\,$cm, several times below the equilibrium-inferred value, indicating that the plasma never relaxes to a pressure-balanced equilibrium within the microsecond pulse. The model further reproduces a non-axisymmetric, four-fold ($m=4$) deformation of the compressed column, matching the square cross-section recorded by the experiment's end-on framing camera, a feature beyond the reach of the two-dimensional models previously applied to this device. Running on modest GPU hardware, this work brings integrated, first-principles kinetic modeling of fusion-relevant FRCs within reach.
\end{abstract}

\maketitle

\section{Introduction}

The field-reversed configuration (FRC) is a compact toroid with negligible toroidal field and very high $\beta\sim1$, which makes it attractive both as a magnetized plasma target for magneto-inertial fusion and as a standalone fusion concept compatible with advanced aneutronic fuels~\cite{tuszewski1988field,steinhauer2011review,pan2023new,wang2025magnetic}. In the classical $\theta$-pinch route, a quartz tube is pre-filled with neutral gas in a bias field, a fast pre-ionization pulse breaks down and magnetizes the gas, and a reversed main-bank current then drives radial compression and field-line reconnection at the tube ends to produce closed flux surfaces~\cite{tuszewski1988field,taccetti2003frx}. Modern devices such as FRX-L, Yingguang-1 and the C-2 family have demonstrated keV-class, $10^{16}$--$10^{17}\,\mathrm{cm^{-3}}$ FRC targets formed on microsecond timescales~\cite{intrator2004high,dongfan2017development,li2016formation,binderbauer2010dynamic}, motivating predictive simulations of the formation phase.

To date, whole-device modeling of FRC formation has relied heavily on resistive or two-temperature magnetohydrodynamics (MHD)~\cite{milroy2010extended,belova2000numerical,li2014two} and, more recently, hybrid models~\cite{winske1996nonspecialist,lipatov2002hybrid}. While MHD captures the gross macroscopic dynamics, it carries severe structural limitations during the highly dynamic formation phase. The driving coil currents are typically prescribed as external boundary conditions without resolving the self-consistent plasma back-reaction, and the fluid approximation entirely misses critical ion finite-Larmor-radius (FLR) effects~\cite{belova2000numerical,steinhauer2011review,zhang2026asymmetric,yao2022detailed}. Crucially, MHD relies on macroscopic closure conditions, such as a fluid Ohm's law, to model magnetic reconnection; because classical resistive reconnection is far too slow to explain sub-microsecond formation timescales, an ad hoc anomalous resistivity must be invoked by hand to approximate the fast reconnection rates observed in experiments~\cite{yamada2010magnetic,yu2023electron,liu2025reconnection}. Hybrid codes, which advance ions as kinetic particles while treating electrons as a massless fluid, resolve the ion-scale physics and have successfully clarified FRC stability and merging~\cite{belova2000numerical,belova2025hybrid}. However, their fluid electron closure remains a fundamental limitation: taking the electron mass to zero removes the electron inertia and electron-scale dynamics that set the electron heating and the self-consistent structure of the reconnecting current layer, the physics that produces the electron-dominated core characteristic of kinetic formation.

A fully kinetic particle-in-cell (PIC) treatment removes this electron fluid approximation, advancing both electrons and ions as discrete particles. Concurrently, an explicit finite-difference time-domain (FDTD) architecture allows the macroscopic drive coils to be embedded directly into the computational grid as physical conductors. This enables the coil currents, the induced fields, and the plasma response to be advanced self-consistently in a single, unified loop without relying on artificial boundary conditions or fluid closures. Until recently, however, such fully kinetic treatments have been considered computationally prohibitive for the formation problem: the device volume is tens of centimeters in each direction, the formation time spans several microseconds, and the explicit CFL condition and Debye constraints push the required number of cell--steps far beyond what CPU PIC codes can deliver.

Two recent developments now make this calculation tractable. First, the GPU systems now built at scale for AI pair large on-package memory and high interconnect bandwidth with exactly the floating-point throughput that the dense, regular kernels of explicit PIC exploit. Second, GPU-native PIC codes such as WarpX~\cite{vay2018warp,fedeli2022pushing} exploit this hardware to deliver orders of magnitude more particle pushes per second per node than legacy CPU codes, putting multi-billion-particle, microsecond-scale runs within reach of a single multi-GPU job. It is this convergence of AI-driven hardware and GPU-native algorithms that turns the whole-device, fully kinetic treatment of FRC formation from a computational aspiration into a tractable problem. The physical payoff is access to a cross-scale regime that has so far remained out of reach: within a single device-scale calculation, the kinetic processes operating at electron and ion scales and the macroscopic compression that drives them can be followed together and self-consistently. The formation phase thus becomes open to direct kinetic study for the first time.

In this paper we present a 3D, whole-device, fully kinetic model built on a GPU-native PIC core, where the macroscopic drive coils are embedded directly as physical conductors and coupled self-consistently to an external circuit. We first detail the numerical methodology, demonstrating how reduced-parameter scaling and the explicit FDTD architecture make the calculation feasible on current multi-GPU hardware. We then validate the method on a closed-loop test configuration, verifying the genuine two-way coupling between the driving $RLC$ circuit, the embedded coils, and the plasma back-reaction. Following this methodological validation, we apply the framework to FRC formation under conditions representative of the Yingguang-1 experiment~\cite{dongfan2017development}. The kinetic baseline reproduces the full formation sequence and the measured peak density, while indicating that the plasma does not relax to the pressure-balanced equilibrium conventionally assumed to infer device separatrix radii and temperatures. Furthermore, the model spontaneously recovers a non-axisymmetric, four-fold ($m=4$) deformation of the compressed core that matches the device's end-on framing-camera images, a genuinely 3D kinetic feature inaccessible to axisymmetric fluid models. Together, these results establish both a robust computational framework and a kinetic reference point for FRC formation that were numerically inaccessible prior to the present generation of GPU PIC tools.

\section{Computational model}

\subsection{Explicit PIC architecture and scaling constraints}

A fully kinetic simulation advances both ions and electrons as particles, and is often assumed to be confined to microscopic domains because the grid must resolve the Debye length $\lambda_{De}$, of order a micron for the Yingguang-1 target, seemingly ruling out a device tens of centimeters across. This bound is not fundamental. Both implicit and explicit schemes lift it and run stably at $\Delta x\gg\lambda_{De}$: implicit schemes by evaluating the fields and particle push at the advanced time level, which damps the under-resolved modes that drive the finite-grid instability~\cite{mason1981implicit,brackbill1982implicit,chen2011energy,lapenta2017exactly}; the explicit scheme by controlling the residual grid heating with high-order particle shape functions and current smoothing~\cite{birdsall2018plasma}. The explicit scheme is then governed by only two constraints, the electron plasma oscillation and the electromagnetic CFL condition,
\begin{equation}
\Delta t\lesssim2/\omega_{pe},\qquad c\,\Delta t/\Delta x \lesssim 1.
\end{equation}

The scale that must genuinely be resolved is therefore the electron skin depth,
\begin{equation}
d_e = \frac{c}{\omega_{pe}} = \sqrt{\frac{\varepsilon_0 m_e c^2}{n_e e^2}},
\end{equation}
which sets the width of the current layers and reconnection sites.

The choice of resolution should therefore be matched to the physics one aims to capture. In a device-scale study the quantities of interest are magnetic field generation, flux compression, ion and electron bulk transport, and magnetic reconnection, all of which are governed by the electron skin depth. The sub-Debye physics averages out without feeding back on these, so the grid is matched to $d_e$.

We adopt the explicit solver, a Yee finite-difference time-domain (FDTD) scheme. This choice is fundamentally motivated by the architecture of modern exascale GPU computing resources. While implicit schemes such as the $\theta$-implicit method are highly attractive for their ability to relax both $\Delta x$ and $\Delta t$, they carry a severe computational bottleneck for device-scale runs. Implicit methods require a global linear solver for the fields, which must be iteratively coupled to the particle push at every timestep. This global, iterative requirement scales poorly to domains with hundreds of millions of grid cells and macroparticles. More critically, global matrix solves map awkwardly onto GPU hardware, which is optimized for massive numbers of simple, localized, and highly regular parallel tasks. The explicit update, by contrast, is entirely local and matrix-free: advancing the fields at a cell requires only its immediate neighbors, and pushing a particle requires only the local fields. This strict locality minimizes memory latency, avoids global reductions, and allows the explicit scheme to saturate the floating-point throughput of modern multi-GPU systems~\cite{vay2018warp,fedeli2022pushing}.

Even at skin-depth resolution the cost is severe for a dense plasma. Since $d_e\propto n_e^{-1/2}$, a $10\times$ density increase shrinks the cell and, in three dimensions, raises the number of cells (the cost of a single step) by about $30\times$. To relieve this, we employ two well-established reduced parameters~\cite{totorica2025particle}:
\begin{itemize}
    \item \textbf{Increased electron mass (reduced mass ratio):} Raising the electron mass by a factor $\alpha$ inflates the skin depth as $d_e\propto\sqrt{\alpha}$ and lengthens the step as $\Delta t\propto\sqrt{\alpha}$. Because the physical electron mass is so small, an artificially increased mass still preserves the vast separation of scales between electrons and ions ($m_e \ll m_i$).
    \item \textbf{Reduced speed of light:} Reducing $c$ further relaxes the Courant condition, which is permissible since both the bulk plasma motion and the particle thermal speeds remain far sub-luminal. We reduce $c$ by raising the vacuum permittivity $\varepsilon_0$, so a tenfold-lower $c$ pairs with a hundredfold-higher $\varepsilon_0$; the product $\varepsilon_0 c^2$, and with it the skin depth $d_e=\sqrt{\varepsilon_0 m_e c^2/n_e e^2}$ that sets the current-layer width, is left untouched.
\end{itemize}
Introducing such artificial parameters is sometimes objected to as unphysical, but every reduced model makes an approximation of the same kind: MHD collapses the plasma to a single fluid closed by an Ohm's law with an ad hoc resistivity and assumes a Maxwellian distribution, and the hybrid model treats the electrons as a massless fluid. The distinction is that the present two knobs are continuously adjustable toward their physical values, not an irreversible assumption but a controlled approximation that converges to the true mass ratio and light speed as resources allow.

In the explicit realization used here, particles are advanced with the Higuera--Cary pusher~\cite{higuera2017structure} and their current is deposited by a charge-conserving Villasenor--Buneman scheme~\cite{villasenor1992rigorous}, while the residual finite-grid heating is held down by third-order (cubic) particle shape functions and two passes of bilinear current smoothing. The outer faces of the domain are terminated by perfectly matched layers, which absorb outgoing electromagnetic radiation, while plasma particles are absorbed at the material walls of the device.

Coulomb collisions are retained between every species pair, electron--electron, ion--ion and electron--ion, through the pairwise binary Monte-Carlo model of Takizuka and Abe~\cite{takizuka1977binary} in its relativistic implementation~\cite{perez2012improved}. Because the mass-inflation factor $\alpha$ rescales the electron collision frequency, the Coulomb logarithm of the two electron-bearing pairs is multiplied by $\sqrt{\alpha}$ to recover the physical collisionality. The plasma thereby evolves from collisional at the cold fill to weakly collisional in the compressed keV core, a transition the binary operator captures self-consistently rather than through a prescribed anomalous resistivity.

\subsection{Embedded coils and external circuit coupling}
\label{sec:coils}

\begin{figure}[htbp]
\centering
\includegraphics[width=0.48\textwidth]{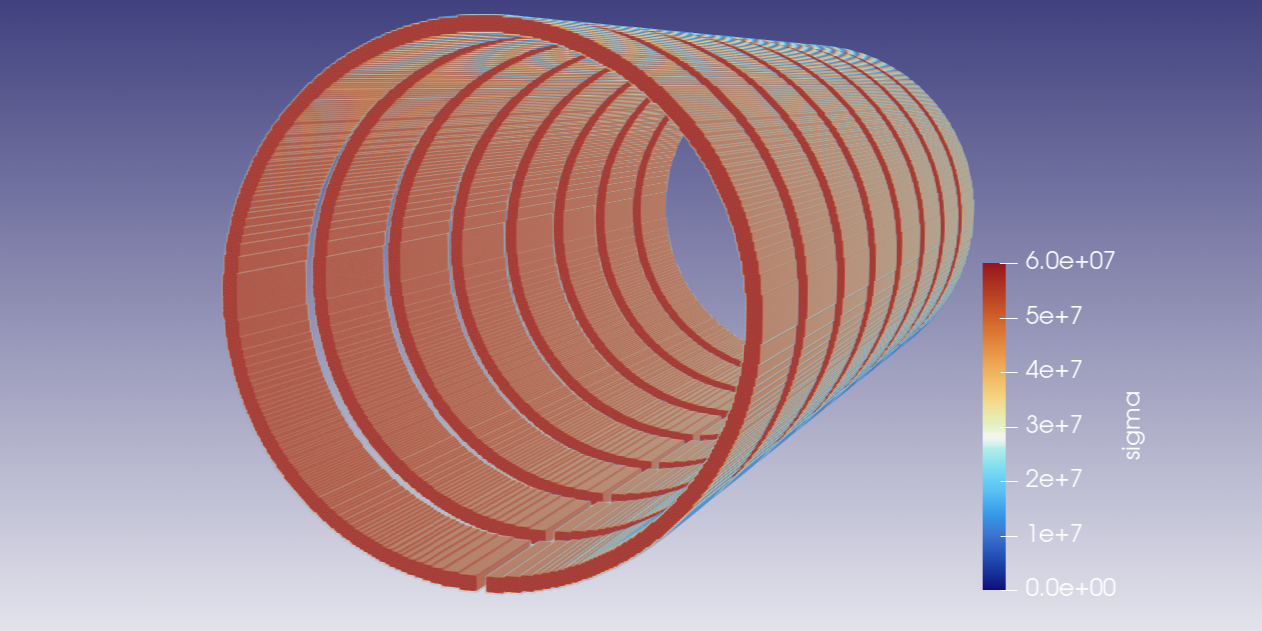}\\[-0.3mm]
{\footnotesize\textbf{(a)}}\par\vskip 1.8mm
\resizebox{0.5\textwidth}{!}{%
\begin{tikzpicture}[x=0.75cm,y=0.75cm,>=Stealth,line width=0.9pt,font=\normalsize]
  \draw[rounded corners=3pt, fill=blue!3, draw=black!55] (0,0) rectangle (9.0,6.3);
  \node[anchor=north west, black!65] at (0.12,6.2) {Kinetic PIC domain};
  \draw[->,black!55,line width=0.7pt] (0.4,0.4)--(1.3,0.4) node[right,inner sep=1pt]{$y$};
  \draw[->,black!55,line width=0.7pt] (0.4,0.4)--(0.4,1.3) node[above,inner sep=1pt]{$x$};
  \def\cu{orange!80!black}
  \draw[\cu, line width=8pt, line cap=round, rounded corners=10pt]
        (7.4,5.0) -- (1.25,5.0) -- (1.25,1.0) -- (7.4,1.0);
  \draw[->,white,line width=1.1pt] (5.3,5.0)--(3.1,5.0);
  \draw[->,white,line width=1.1pt] (1.25,3.6)--(1.25,2.4);
  \draw[->,white,line width=1.1pt] (3.1,1.0)--(5.3,1.0);
  \node[orange!45!black,rotate=90,anchor=center] at (0.6,3.3) {Cu coil $\sigma(\mathbf{x})$};
  \draw[fill=red!12,draw=red!55] (4.1,3.0) ellipse (2.0 and 1.15);
  \draw[red!45] (4.1,3.0) ellipse (1.28 and 0.72);
  \draw[red!45] (4.1,3.0) ellipse (0.6 and 0.34);
  \node[red!55!black] at (4.1,3.0) {FRC};
  \coordinate (Tp) at (7.4,5.0);
  \coordinate (Bp) at (7.4,1.0);
  \draw[dashed,black!60,line width=0.7pt] (7.4,4.7)--(7.4,1.3);
  \draw[<->,black!70,line width=0.7pt] (7.75,4.6)--(7.75,1.4);
  \node[draw=black!55,fill=white,rounded corners=2pt,inner sep=2.5pt,anchor=west]
        at (8.05,3.0) {$V_{\rm port}$};
  \coordinate (Ct) at (10.5,5.0);
  \coordinate (Cb) at (10.5,1.0);
  \draw (Tp)--(Ct);
  \draw (Bp)--(Cb);
  \draw (Ct)--(10.5,4.25);
  \draw[decorate,decoration={zigzag,segment length=6pt,amplitude=4pt}] (10.5,4.25)--(10.5,3.35);
  \node[anchor=west] at (10.85,3.8) {$R$};
  \draw (10.5,3.35)--(10.5,3.1);
  \draw[decorate,decoration={coil,segment length=5pt,amplitude=5pt,aspect=0.6}] (10.5,3.1)--(10.5,2.05);
  \node[anchor=west] at (10.95,2.55) {$L$};
  \draw (10.5,2.05)--(10.5,1.8);
  \draw[line width=1.4pt] (10.1,1.8)--(10.9,1.8);
  \draw[line width=1.4pt] (10.1,1.5)--(10.9,1.5);
  \node[anchor=west] at (10.95,1.65) {$C$};
  \draw (10.5,1.5)--(Cb);
  \draw (10.5,1.0)--(10.5,0.5);
  \draw[line width=0.9pt] (10.0,0.5)--(11.0,0.5);
  \draw[line width=0.9pt] (10.18,0.31)--(10.82,0.31);
  \draw[line width=0.9pt] (10.36,0.14)--(10.64,0.14);
  \node[black!75,rotate=90,anchor=center] at (12.0,3.0) {External circuit};
  \node[draw,rounded corners,fill=black!3,align=center,font=\small,anchor=north]
        at (5.3,-0.25) {open loop $I(t)$\ \ $\Longleftrightarrow$\ \ closed-loop $RLC$ bank};
\end{tikzpicture}}\\[0.5mm]
{\footnotesize\textbf{(b)}}
\caption{\label{fig:coils} Coils carried inside the simulation as physical conductors. (a) The drive coils voxelized onto the Cartesian grid as a spatially varying conductivity $\sigma(\mathbf{x})$ set to the true copper value ($\sigma\!\sim\!6\times10^{7}\,\mathrm{S/m}$), rendered in the actual simulation domain. (b) Self-consistent coil--circuit coupling: the voxelized coil (copper, orange) shares the same FDTD grid as the plasma (FRC, red). The coil is energized across a port gap into which a current density $\mathbf{J}=I(t)/A$ is deposited; the port voltage $V_{\rm port}=\oint\mathbf{E}\cdot d\mathbf{l}$ is read back from the field and either fed to an external series-$RLC$ bank (closed loop) or replaced by a prescribed waveform (open loop). The ground symbol denotes the circuit reference node only.}
\end{figure}

A distinctive feature of the present model is that the drive coils are not imposed as an external boundary field but are carried inside the simulation as physical conductors. Reduced plasma models cannot do this: they advance the field through a plasma Ohm's law that has no meaning inside a solid conductor, so the coils are kept outside the domain and their field prescribed as an external source. That convenience discards the physics of a fast-pinch circuit: the coil self-inductance and mutual coupling, the eddy currents and field diffusion in the copper, and, above all, the back-reaction of the plasma on the drive current. The coil current is then a prescribed input rather than a self-consistent response.

The explicit FDTD field solver removes this artificial separation at almost no extra cost. The same locality that makes the scheme GPU-friendly means a conductor enters the update only through the local Ohmic current $\sigma\mathbf{E}$ in Amp\`ere's law, which the field update folds in implicitly through a backward-Euler coefficient,
\begin{equation}
\mathbf{E}^{n+1}=C_a\,\mathbf{E}^{n}+C_b\left(\nabla\times\mathbf{H}-\mathbf{J}_{\mathrm{src}}\right),
\end{equation}
where $C_a=(1+\sigma\Delta t/\varepsilon)^{-1}$, $C_b = C_a\Delta t/\varepsilon$, and $\mathbf{J}_{\mathrm{src}}$ is any impressed (port-injection) current, the Ohmic response $\sigma\mathbf{E}$ being absorbed into $C_a$; the scheme thus stays unconditionally stable even at the very large conductivity of copper. Modelling a coil therefore reduces to embedding a region of high $\sigma$ in the grid; no separate conductor solver, and no global field solve, is required.

In practice each coil is voxelized onto the Cartesian mesh as a spatially varying conductivity $\sigma(\mathbf{x})$ set to the true copper value, so that all of these effects, together with the self-consistent return-current path, emerge from the same Maxwell update that advances the plasma fields. The coil is energized across a thin gap, or port, into which a current density $\mathbf{J}=I(t)/A$ is deposited each step (Fig.~\ref{fig:coils}). Two interchangeable backends drive this port. In the \emph{open-loop} mode a tabulated waveform $I(t)$, taken from an external bank model or the measured experimental current, is interpolated onto the timestep, keeping the code agnostic to the details of the source. In the \emph{closed-loop} mode a lumped series-$RLC$ bank is advanced in lockstep with the field solve: the port voltage is measured directly from the field as the line integral $V_{\rm port}=\oint\mathbf{E}\cdot d\mathbf{l}$ across the gap, and the branch current is advanced by an implicit (backward-Euler) update of the series-$RLC$ equation, with the capacitor voltage and resistive losses tracked alongside, closing a self-consistent coil--circuit--plasma loop.

\begin{figure*}[t]
\centering
\includegraphics[width=0.7\textwidth]{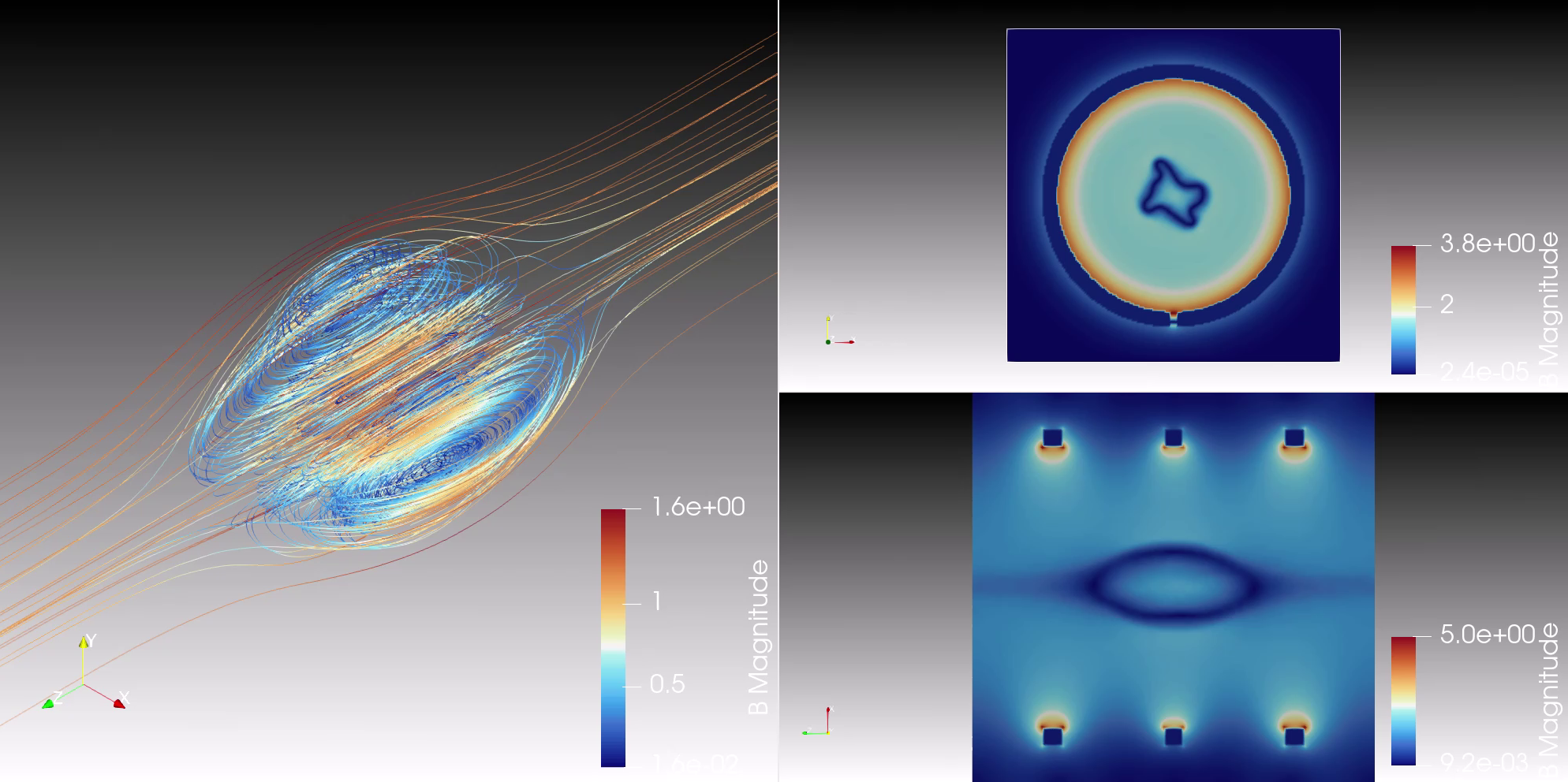}\\[0.3mm]
{\footnotesize\textbf{(a)}}\par\vskip 1.2mm
\includegraphics[width=0.7\textwidth]{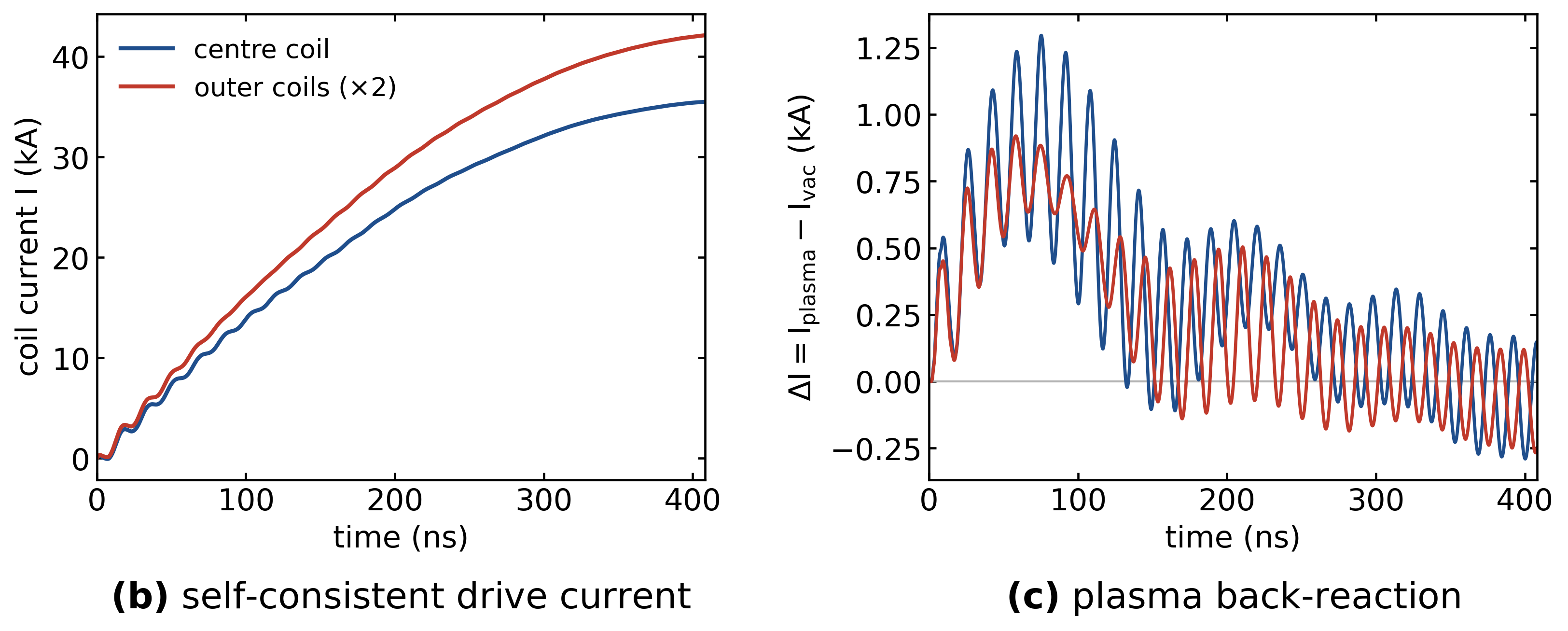}
\caption{\label{fig:rlc} Closed-loop coil--circuit--plasma coupling, demonstrated on a three-coil test configuration in which each coil bank is driven by its own series-$RLC$ branch ($R=0.5\,\Omega$, $L=255\,$nH, $C=230\,$nF, charged to $62$ and $74.4\,$kV); the branch current is advanced solely from the field-measured port voltage $V_{\rm port}=\oint\mathbf{E}\cdot d\mathbf{l}$. (a) Spatial state of the run: the 3D magnetic field lines colored by $|\mathbf{B}|$, a transverse $|\mathbf{B}|$ slice (upper right), and an axial slice (lower right) showing the three voxelized coils (dark) with the plasma column between them. (b) Self-consistent coil currents: the branch current is an output of the field-coupled $RLC$ solve, ringing up to ${\approx}35\,$kA (centre) and ${\approx}42\,$kA (outer coils). (c) Plasma back-reaction, isolated as the difference $\Delta I=I_{\mathrm{plasma}}-I_{\mathrm{vac}}$ between the run and an identical vacuum run: the plasma loads the circuit and shifts the self-consistent current by up to ${\approx}1.3\,$kA.}
\end{figure*}

We verify this backend on a reduced three-coil configuration in which each coil is energized only through its own series-$RLC$ branch ($R=0.5\,\Omega$, $L=255\,$nH, $C=230\,$nF), the centre bank charged to $62\,$kV and the outer pair to $74.4\,$kV, the branch current being advanced solely from the field-measured port voltage with no prescribed waveform anywhere in the loop (Fig.~\ref{fig:rlc}). The current emerges as a genuine output of the coupled solve, ringing up as the banks discharge to ${\approx}35\,$kA in the centre coil and, driven by its higher charge, ${\approx}42\,$kA in the outer pair, while the port voltage rings on the fast circuit timescale and damps as energy is delivered to the plasma. That the coupling is truly two-way is shown by repeating the run in vacuum: the plasma loads the circuit and shifts the self-consistent current by up to ${\approx}1.3\,$kA, a few percent of the peak, a back-reaction that no prescribed-current source can reproduce. The two timescales in Fig.~\ref{fig:rlc} have distinct origins: the slow rise of the current over ${\sim}0.4\,\mu$s is the quarter-period of the lumped bank's own $LC$ resonance, whereas the fast ${\sim}60\,$MHz oscillation carried by the port voltage, nearly two orders of magnitude higher, is set by the self- and mutual-inductance of the embedded coils together with the stray capacitance of the drive gap. Because the coils are advanced as physical conductors rather than lumped branch elements, this geometric resonance, and the inter-coil coupling that shifts it, emerge directly from the field solution rather than being supplied by hand.

\section{Simulation of the Yingguang-1 FRC formation}

\subsection{Model setup and initial conditions}

\begin{figure}[htbp]
\centering
\resizebox{0.47\textwidth}{!}{%
\begin{tikzpicture}[x=0.13cm,y=0.13cm,>=Stealth,line width=0.6pt,font=\footnotesize]
  \foreach \rr in {-2.6,2.6}{\draw[->,blue!35,line width=0.4pt] (-15.5,\rr)--(15.5,\rr);}
  \node[blue!45!black,anchor=south] at (-11,2.7) {$B_z$};
  \draw[rounded corners=2pt,fill=blue!4,draw=black!50] (-19.2,-8) rectangle (19.2,8);
  \node[anchor=south,black!60] at (0,8.4) {kinetic PIC domain ($16\times38.4$\,cm)};
  \fill[black!18] (-18.5,5.25) rectangle (18.5,5.5);
  \fill[black!18] (-18.5,-5.5) rectangle (18.5,-5.25);
  \fill[red!12] (-18,-5.15) rectangle (18,5.15);
  \node[red!55!black,align=center] at (0,0) {initial flash-\\injection region};
  \foreach \zc in {-15.75,-11.25,-6.75,-2.25,2.25,6.75,11.25,15.75}{
    \fill[orange!80!black] (\zc-1.75,6.2) rectangle (\zc+1.75,6.6);
    \fill[orange!80!black] (\zc-1.75,-6.6) rectangle (\zc+1.75,-6.2);
  }
  \foreach \zm in {-25.5,25.5}{
    \fill[Mahogany] (\zm-1.0,8.0) rectangle (\zm+1.0,9.6);
    \fill[Mahogany] (\zm-1.0,-9.6) rectangle (\zm+1.0,-8.0);
  }
  \node[Mahogany,anchor=south,align=center] at (25.5,9.8) {mirror coil\\(analytic)};
  \draw[black!55,line width=0.3pt] (6.75,-6.65)--(8,-9.4);
  \node[orange!50!black,anchor=north] at (8,-9.5) {Cu $\theta$-coils ($\times8$)};
  \draw[black!55,line width=0.3pt] (-10,-5.42)--(-10,-9.4);
  \node[black!55,anchor=north] at (-10,-9.5) {quartz tube};
  \draw[->,black!60,line width=0.5pt] (-28,-11)--(-23,-11) node[right,inner sep=1pt]{$z$};
  \draw[->,black!60,line width=0.5pt] (-28,-11)--(-28,-7.4) node[above,inner sep=1pt]{$x$};
\end{tikzpicture}}\\[0.5mm]
{\footnotesize\textbf{(a)}}\par\vskip 1.8mm
\includegraphics[width=0.38\textwidth]{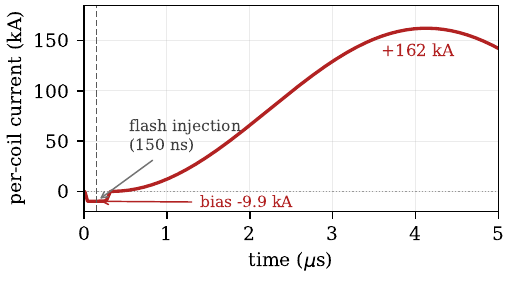}\\[0.5mm]
{\footnotesize\textbf{(b)}}
\caption{\label{fig:device} Whole-device model of Yingguang-1 formation. (a) Axial cross-section of the modeled device: the eight copper $\theta$-pinch coils (orange), carried on the grid as a spatially varying conductivity $\sigma(\mathbf{x})$; the two end-mirror coils imposed analytically outside the grid (dark red); the quartz tube (grey); and the flash-injected deuterium column (red, the initial plasma region). (b) Prescribed per-coil drive waveform $I(t)$.}
\end{figure}
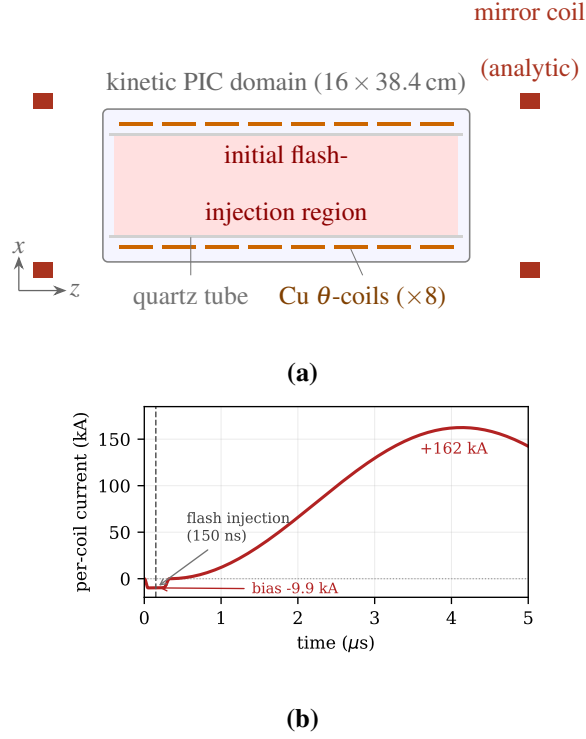

We apply the framework to a three-dimensional, whole-device model of the Yingguang-1 formation experiment~\cite{dongfan2017development}; the model inputs and numerical parameters are collected in Table~\ref{tab:config}. The eight inner theta-pinch coils are carried as physical copper conductors ($\sigma=6\times10^{7}\,\mathrm{S/m}$) on the same grid as the plasma, stacked on a 4.5-cm axial pitch to reproduce the as-built array. To directly reproduce the experimental drive history, the coils are energized in the open-loop mode through a two-cell port gap; this approach is necessary because the published device parameters are insufficient to properly constrain a full closed-loop bank model. The $16\times16\times38.4\,\mathrm{cm^{3}}$ domain is resolved at uniform $0.5\,$mm spacing; with the reduced-parameter scaling introduced above, the full $5\,\mu$s discharge is reached in ${\sim}5\times10^{5}$ steps.

The two end-mirror solenoids are not resolved as conductors: their quarter-period (${\sim}700\,\mu$s) far exceeds the discharge, so the mirror field is effectively static and is imposed as an analytic axial bias from two 20-turn loops at $z=\pm25.5\,$cm carrying $7\,$kA. The fused-quartz confinement tube ($\varepsilon_r=3.8$, nonmagnetic) is embedded as a dielectric annulus whose inner wall absorbs particles.

The pre-ionized target is introduced at $t=150\,$ns by flash injection of a fully ionized, quasi-neutral deuterium column ($T_0=2\,$eV, radius $5.15\,$cm, length $36\,$cm), representing the plasma state at the end of the ionization stage; electron--electron, deuteron--deuteron and electron--deuteron Coulomb collisions are retained. The prescribed coil current (Fig.~\ref{fig:device}(b)) follows the programmed Yingguang-1 sequence: a slowly rising reversed bias is established and held, then crosses zero and rises as a $\sin^{2}$ main theta-pinch pulse ($1.3\,$MA total over the eight coils), matching the measured per-coil waveform.

\begin{table*}[t]
\caption{\label{tab:config}Configuration of the whole-device model, matched to the as-built Yingguang-1 hardware~\cite{dongfan2017development}, and the numerical simulation parameters.}
\setlength{\tabcolsep}{6pt}
\begin{ruledtabular}
\begin{tabular}{lll}
Quantity & This work & Yingguang-1 \\
\hline
\multicolumn{3}{l}{\textit{Configuration (matched to device)}}\\
Coil inner diameter            & 12.4\,cm                       & 12.4\,cm \\
Coil count $\times$ width      & $8\times3.5$\,cm               & $8\times3.5$\,cm \\
Active coil length             & 36\,cm                         & 36\,cm \\
Quartz radii (in/out)          & 5.25 / 5.5\,cm                 & 5.25 / 5.5\,cm \\
Per-coil bias current          & $-9.9$\,kA                     & ${\sim}10$\,kA \\
Per-coil $\theta$-pinch        & $162$\,kA                      & ${\sim}160$\,kA \\
$\theta$-quarter-period        & $3.8\,\mu$s                    & $3.8\,\mu$s \\
Fill density $n_0$             & $2\times10^{15}\,\mathrm{cm^{-3}}$ & ${\sim}2\times10^{15}\,\mathrm{cm^{-3}}$ \\
Working ion                    & D                              & H/D/Ar \\
\hline
\multicolumn{3}{l}{\textit{Numerical parameters (this work)}}\\
Mass inflation $\alpha$        & \multicolumn{2}{l}{15} \\
Speed of light scaling         & \multicolumn{2}{l}{$c/10$} \\
Particles/cell/species         & \multicolumn{2}{l}{4} \\
Total macroparticles           & \multicolumn{2}{l}{${\sim}$190M} \\
Cell size $\Delta x,\Delta y,\Delta z$ & \multicolumn{2}{l}{0.5\,mm} \\
Timestep $\Delta t$            & \multicolumn{2}{l}{$9.5\times10^{-12}$\,s} \\
Total cell count               & \multicolumn{2}{l}{79M} \\
\end{tabular}
\end{ruledtabular}
\end{table*}

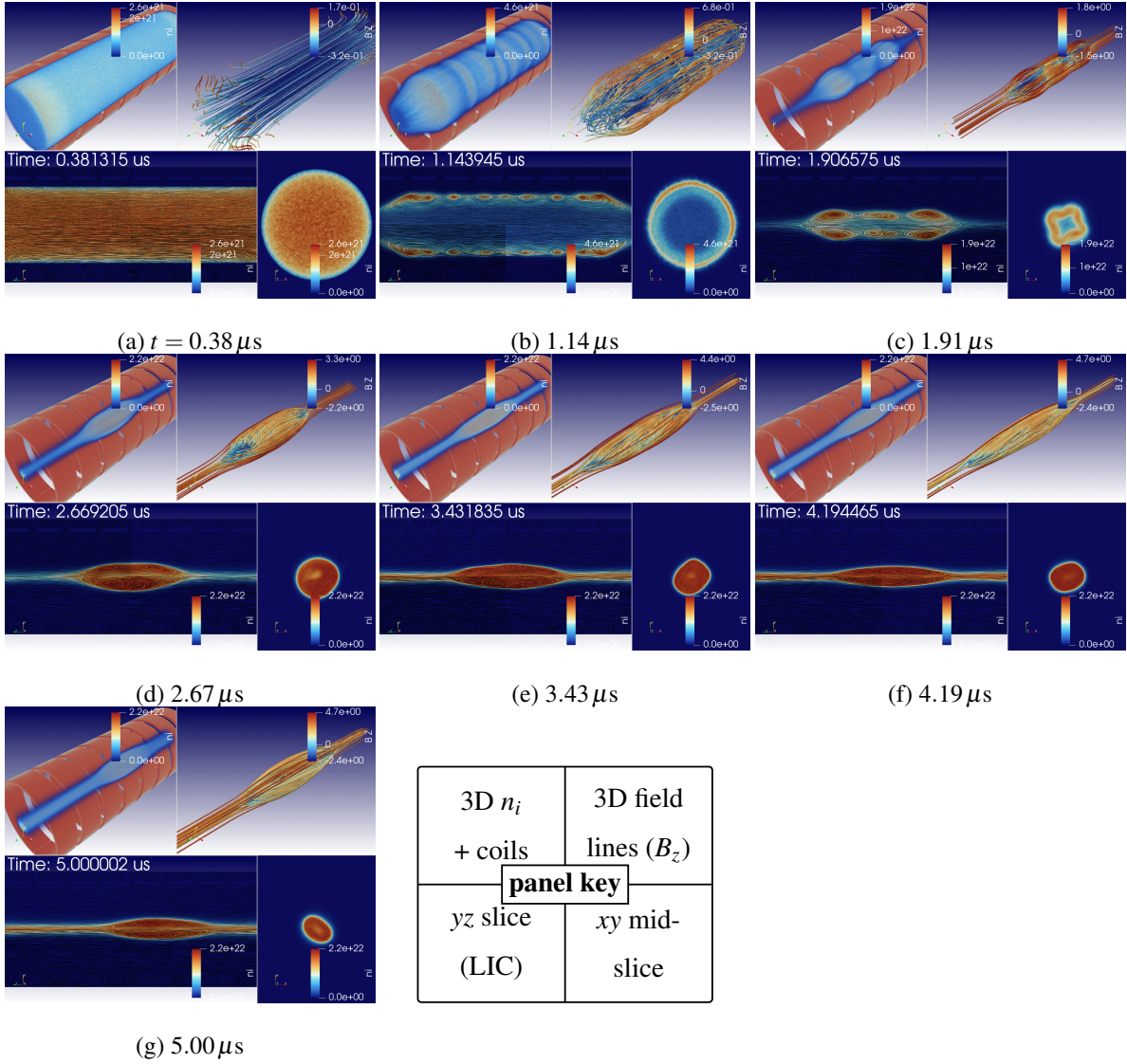
\begin{figure*}[t]
\centering
\setlength{\tabcolsep}{1pt}
\def\frm#1#2{\begin{minipage}[t]{0.32\textwidth}\centering\includegraphics[width=\linewidth]{{#1}.png}\\[-0.4mm]{\footnotesize #2}\end{minipage}}%
\resizebox{0.95\textwidth}{!}{%
\begin{tabular}{@{}ccc@{}}
\frm{prod_0.38}{(a) $t=0.38\,\mu$s} & \frm{prod_1.14}{(b) $1.14\,\mu$s} & \frm{prod_1.9}{(c) $1.91\,\mu$s} \\[1.2mm]
\frm{prod_2.66}{(d) $2.67\,\mu$s} & \frm{prod_3.43}{(e) $3.43\,\mu$s} & \frm{prod_4.19}{(f) $4.19\,\mu$s} \\[1.2mm]
\frm{prod_5}{(g) $5.00\,\mu$s} &
\begin{minipage}[t]{0.32\textwidth}\centering
\resizebox{0.8\linewidth}{!}{%
\begin{tikzpicture}[font=\scriptsize,line width=0.5pt]
\draw[rounded corners=1pt] (0,0) rectangle (3.0,2.4);
\draw (1.5,0)--(1.5,2.4);
\draw (0,1.2)--(3.0,1.2);
\node[align=center] at (0.75,1.8){3D $n_i$\\+ coils};
\node[align=center] at (2.25,1.8){3D field\\lines ($B_z$)};
\node[align=center] at (0.75,0.6){$yz$ slice\\(LIC)};
\node[align=center] at (2.25,0.6){$xy$ mid-\\slice};
\node[draw, fill=white, inner sep=2pt, font=\scriptsize\bfseries] at (1.5,1.2) {panel key};
\end{tikzpicture}%
}
\end{minipage} \\
\end{tabular}}
\caption{\label{fig:sequence} Time sequence of field-reversed-configuration formation in the whole-device kinetic simulation. (a)--(g) Snapshots at $t=0.38$, 1.14, 1.91, 2.67, 3.43, 4.19 and $5.00\,\mu$s; each frame shows four synchronized views (panel key, lower right): the 3D ion density with the copper coils (orange), the 3D magnetic field lines colored by $B_z$, an axial $yz$ density slice with the in-plane field traced by line-integral convolution, and a transverse $xy$ density slice at the midplane. The core compresses to a peak ion density ${\sim}2.2\times10^{22}\,\mathrm{m^{-3}}$.}
\end{figure*}

\subsection{Magnetic reconnection and FRC formation sequence}

The whole-device run reproduces the complete $\theta$-pinch formation sequence (Fig.~\ref{fig:sequence}); each frame combines the three-dimensional ion density and magnetic field lines with an axial ($yz$) density slice, on which the in-plane field is traced by line-integral convolution (LIC), and a transverse ($xy$) slice at the midplane. At $t=0.38\,\mu$s the reversed bias field is frozen into the pre-ionized column as a nearly uniform axial flux. As the main pinch current rises the plasma diamagnetically expels the incoming field, and by $t=1.14\,\mu$s the induced currents have organized into a chain of seven azimuthal eddy cells, three on each side of the midplane and one at the centre, while the field lines begin to wind helically about the axis. These cells are the seeds of reconnection. The first tearing and merging is underway by $t=1.91\,\mu$s, where the bias and main flux reconnect and the seven cells coalesce into one lobe on each side and two at the centre, closing the first poloidal flux surfaces. Reconnection here is not a single relaxation event, however, but a recurring one: as the main pinch continues to drive freshly compressed flux into the contracting current sheet, the layer repeatedly reforms and tears, so that the configuration passes through several successive rounds of reconnection between ${\approx}2$ and ${\approx}4.5\,\mu$s rather than reconnecting once and settling. A closed-flux FRC is nonetheless established by $t=2.67\,\mu$s, an elongated dense core bounded by a clear separatrix with the traced field lines retaining a pronounced axial rotation; from $t=3.43$ to $5\,\mu$s this core is long-lived, persisting to the end of the discharge with the field lines gradually straightening, yet it continues to reorganize internally through ongoing reconnection rather than relaxing to a static, pressure-balanced equilibrium. This repetitive reconnection is the kinetic engine behind the sustained electron heating discussed below: because the current layer is continually reformed and reconnected across this window, magnetic energy is funneled into the electrons at the layer throughout, rather than released in a single burst, holding $T_e$ above $T_i$ from ${\approx}2$ to ${\approx}4.5\,\mu$s (Fig.~\ref{fig:kinetic}). The compressed core reaches a peak ion density of ${\sim}2.2\times10^{22}\,\mathrm{m^{-3}}$ ($2.2\times10^{16}\,\mathrm{cm^{-3}}$), consistent with the density reported for Yingguang-1~\cite{sun2017formation}.

\subsection{Azimuthal symmetry breaking and comparison with end-on imaging}
\label{sec:endon}

\begin{figure*}[t]
\centering
\includegraphics[width=0.7\textwidth]{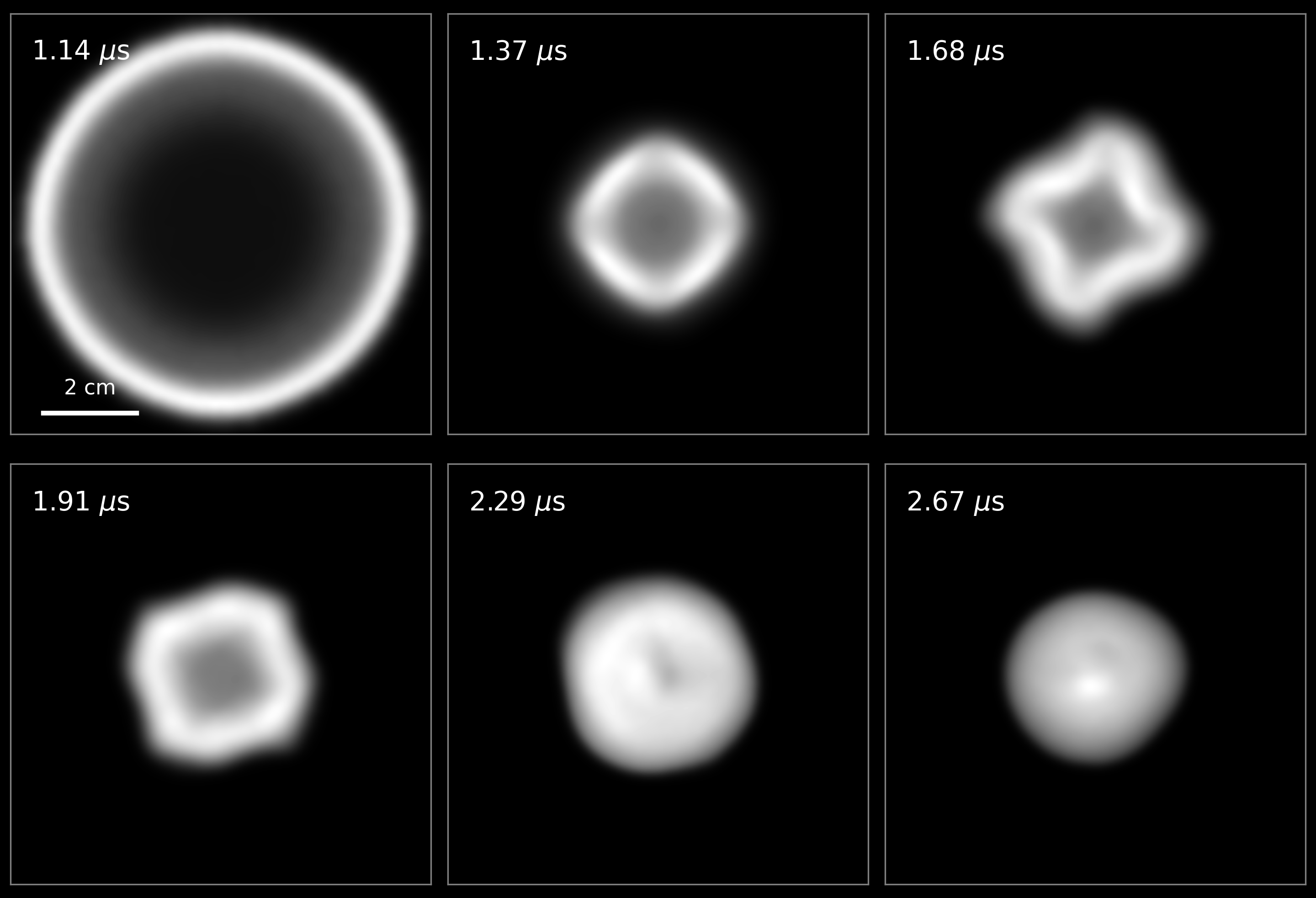}
\caption{\label{fig:endon} Synthetic end-on images of the imploding core, each formed by integrating the emission proxy $\int n_e^2\,dz$ along the axial line of sight to mimic the recombination-dominated self-luminance recorded by the experimental framing camera, shown through the implosion at $t\approx1.14$, 1.37, 1.68, 1.91, 2.29 and $2.67\,\mu$s. Intensity is in arbitrary units and each frame is independently normalized, as in the variable-gain framing-camera diagnostic, so the panels convey morphology rather than absolute brightness. An initially circular sheath ($1.14\,\mu$s) develops a pronounced four-fold ($m=4$), square cross-section (clearest near $1.7$--$1.9\,\mu$s) before compressing into a dense core. The sequence is to be compared with the end-on framing-camera images of the Yingguang-1 experiment, which show the same circle-to-square evolution as the FRC compresses (Ref.~\cite{sun2017formation}, Fig.~15(c); the square is clearest near $t=577\,\mu$s).}
\end{figure*}

Sectioned in the transverse ($xy$) plane, the field-reversal layer is not azimuthally symmetric. As the column implodes it first appears as a circular sheath (at $t\approx1.14\,\mu$s) that becomes weakly modulated and then sharpens into a pronounced four-fold, square cross-section, clearest near peak implosion ($t\approx1.7$--$1.9\,\mu$s), before being carried inward as the core compresses ($t\approx2.3$--$2.7\,\mu$s).

This azimuthal symmetry breaking has a direct experimental counterpart. The end-on framing-camera images of the Yingguang-1 formation experiment show the compressed luminous core taking the same polygonal, four-fold shape, rather than a circular ring, in the compressed FRC (Ref.~\cite{sun2017formation}, Fig.~15(c) near $t=577\,\mu$s). To compare with this diagnostic directly we synthesize an end-on image by integrating an emission proxy along the axial line of sight, $I(x,y)=\int n_e^2\,dz$, evaluated from the deuteron density ($n_e=n_i$ for the fully ionized, singly charged deuterium), over the camera field of view (Fig.~\ref{fig:endon}). Crucially, such a mode cannot appear in the axisymmetric two-dimensional MHD modeling used previously, which is azimuthally uniform by construction and reproduces only circular end-on images~\cite{sun2017formation}; capturing it requires a fully three-dimensional treatment. A square ($m=4$) deformation is, moreover, physically expected in this device class: it is the next harmonic of the ion-spin-up--driven rotational mode that governs field-reversed theta-pinch stability~\cite{belova2004kinetic}, and quadrupole- and octupole-field stabilization is known to impose exactly such polygonal equilibria on these plasmas~\cite{ohi1983quadrupole}. Here the mode instead emerges spontaneously and is resolved kinetically during the formation phase itself.

\subsection{Kinetic state of the compressed core}

\begin{figure*}[t]
\centering
\includegraphics[width=0.7\textwidth]{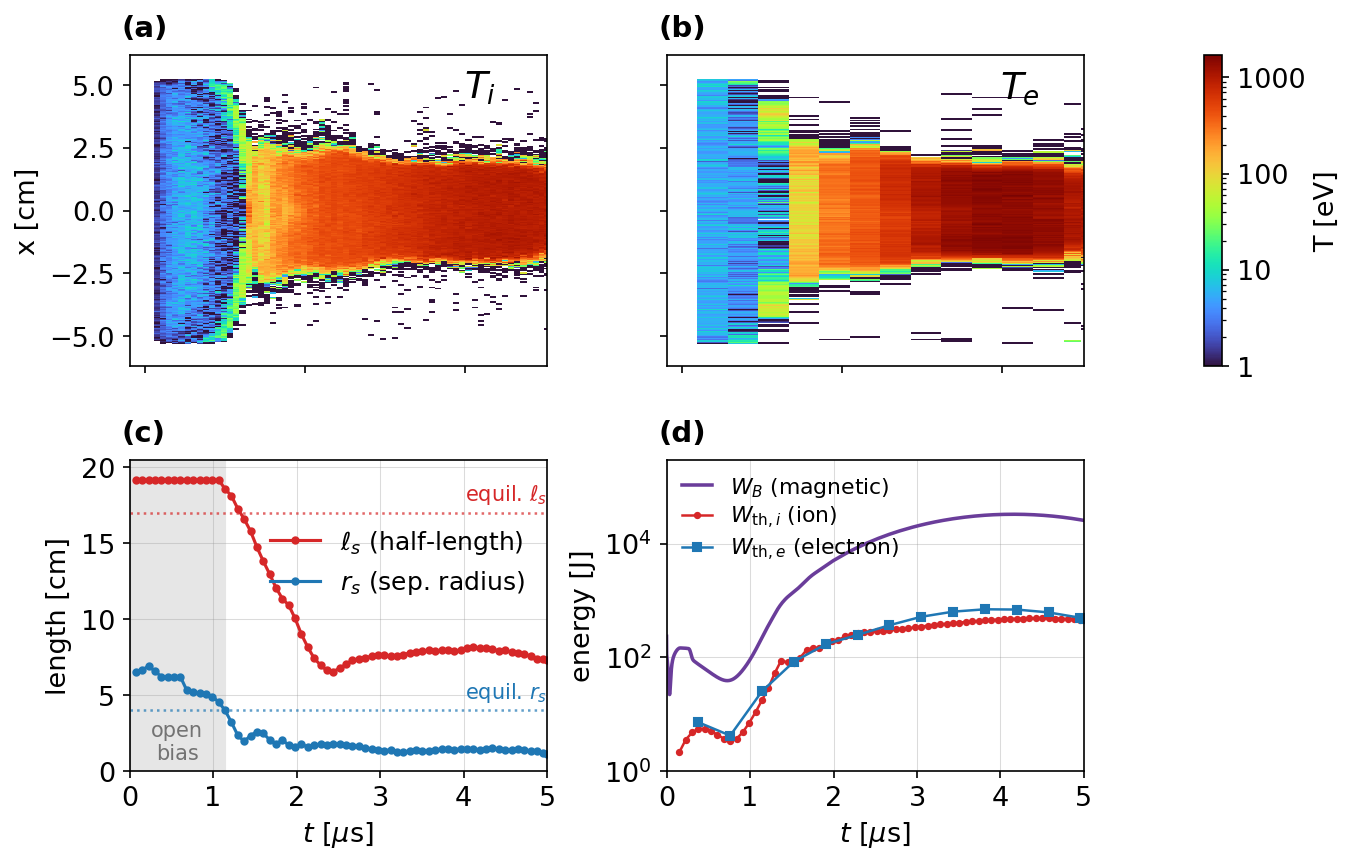}
\caption{\label{fig:kinetic} Kinetic state of the compressed configuration. (a,b) Midplane radius--time streaks of the ion ($T_i$) and electron ($T_e$) temperatures along a transverse cut through the field null; the shared color scale is logarithmic in eV and white marks field-free vacuum (the coarser $T_e$ banding reflects the lower-cadence full diagnostic dump). (c) Separatrix radius $r_s$ and half-length $\ell_s$, extracted frame-by-frame from the axial-field reversal; the shaded band is the pre-formation phase in which the reversed-bias column still spans the device, and the dotted lines mark the equilibrium-inferred device values. (d) Whole-domain energy budget on a logarithmic scale: the magnetic field energy ($W_B$) and the ion and electron thermal energies $W_{\mathrm{th},s}=\tfrac32\!\int n_s k_B T_s\,dV$.}
\end{figure*}

The same run resolves the kinetic state of the compressed plasma (Fig.~\ref{fig:kinetic}). The separatrix radius $r_s(t)$ and half-length $\ell_s(t)$, extracted frame-by-frame from the axial-field reversal (the on-axis $B_z<0$ column sets $\ell_s$; its widest zero-crossing sets $r_s$), trace the compression arc of panel (c). Once the bias flux closes off near $t\approx1.1\,\mu$s the configuration contracts rapidly: $\ell_s$ falls from the device length to ${\approx}7\,$cm and $r_s$ from the bias scale to ${\approx}1.5\,$cm by $t\approx2.4\,\mu$s. It then holds quasi-static for the rest of the pulse, $r_s$ settling to ${\sim}1$--$1.5\,$cm and reaching ${\approx}1\,$cm by the end of the discharge. The contraction is complete well within the quarter-period of the drive, leaving an elongated FRC (elongation $E=\ell_s/r_s\approx6$) markedly smaller than the size reported for the device ($r_s\approx4\,$cm, half-length ${\approx}17\,$cm~\cite{sun2017formation}). The reported values are not measured geometrically but inferred from magnetic- and flux-probe signals through an excluded-flux relation that assumes a pressure-balanced FRC equilibrium. Our fully kinetic result invokes no such closure, indicating that under the strong, high-current pulsed compression of this device the plasma does not relax to that equilibrium within the pulse, so the equilibrium inversion returns a larger size than the plasma reaches within the pulse. This is corroborated by the device's own time-resolved data: the separatrix radii measured during compression fall well below $4\,$cm, approaching the ${\sim}1.4\,$cm reached by the accompanying MHD simulation at peak compression~\cite{sun2017formation}.

The midplane temperature streaks (panels a,b) show both species heated to keV scale during the compression: $T_i$ reaches ${\approx}1.2\,$keV and $T_e\approx1.7\,$keV, concentrated in the contracting core. The electron temperature exceeds the ion temperature throughout ($T_e>T_i$), a distinctly kinetic signature of strong electron heating at the reconnection current layer that is obscured by the fluid-electron approximations of MHD and hybrid models. Both species run several times hotter than the single equilibrium temperature of ${\approx}200\,$eV quoted for the device~\cite{sun2017formation}, which, like the $4\,$cm radius, is not measured but inferred as a total temperature from the same pressure-balance closure; the kinetic run instead resolves the two species separately and finds the more strongly compressed core (smaller $r_s$, higher field) correspondingly hotter, reinforcing that the plasma does not settle to the assumed equilibrium within the pulse. In the whole-domain energy budget (panel d), the pinch drives the magnetic energy from its ${\sim}0.2\,$kJ static floor to ${\sim}30\,$kJ, while the plasma thermal energy grows from a few joules at injection to ${\sim}0.5$--$0.7\,$kJ, the electrons again leading the ions ($W_{\mathrm{th},e}\approx0.7\,$kJ versus $W_{\mathrm{th},i}\approx0.5\,$kJ). The electric-field energy remains negligible throughout, never exceeding a few joules---some five orders of magnitude below $W_B$---confirming that the dynamics are magnetically dominated and quasineutral, so it is omitted from the panel.

\section{Discussion}

The principal approximation that makes this calculation tractable is the reduced-parameter scaling (an electron mass inflated by $\alpha=15$ and a tenfold-reduced speed of light), which relaxes the electron plasma frequency and the explicit CFL limit. The choice is physically sound and leaves the bulk formation dynamics intact, but it may shift the quantities set at electron scales, the inter-species temperature partition most of all. Quantifying that shift, by relaxing the parameters toward their physical values, is an interesting direction for future work, and one that growing computational resources now bring within reach.

The full $5\,\mu$s discharge was reached on a single four-GPU node (NVIDIA V100, $32\,$GB each; $128\,$GB aggregate) in $3.19\times10^{5}\,$s of wall-clock time, roughly $3.7$ days. That this is already feasible on now-obsolete hardware is the essential point, because the trajectory of accelerator memory has since made the remaining approximations optional. A single next-generation rack such as the GB300 NVL72 integrates 72 Blackwell-class GPUs behind $20\,$TB of unified GPU memory at up to $576\,\mathrm{TB/s}$, more than two orders of magnitude beyond the $128\,$GB used here, with comparable gains in bandwidth and throughput. That memory headroom lifts the domain-size constraint, while the bandwidth and compute lift the time-step constraint that lengthening the integration to physical electron mass and light speed would impose. On such a platform the same device could be run from first principles at the true deuteron-to-electron mass ratio and the physical speed of light, or, holding the reduced parameters fixed, the domain could be scaled up toward a reactor-relevant device an order of magnitude larger in linear size. Either way the formation problem moves from a parameter-scaled demonstration to a routine, predictive calculation.

On the algorithmic side the model is being extended beyond formation. A Monte-Carlo nuclear-reaction module for D--D and D--T fusion is in testing, and multi-stage ionization and radiation-transport modules are planned, so that the same self-consistent loop can follow a target from neutral fill through ionization and compression into the burn phase. Whole-device kinetic modeling of an FRC (coils, circuit, plasma, and soon the fusion products themselves) is no longer out of reach; the route to it starts here.

\section{Conclusion}

In conclusion, whole-device, fully kinetic modeling of FRC formation, long thought impractical, is now feasible. Our three-dimensional model carries the electrons as kinetic particles and embeds the drive coils directly in the grid as current-carrying conductors, so that their fields and the plasma response evolve together. Applied to the Yingguang-1 $\theta$-pinch, it reproduces the complete formation sequence through to a closed-flux FRC. Treating the electrons kinetically resolves what fluid and hybrid closures cannot. We find that the configuration does not reconnect once and settle but passes through repeated rounds of tearing and reconnection at the current layer, and that this recurring reconnection is the engine that channels magnetic energy into the electrons and sustains an electron-dominated core ($T_e>T_i$) throughout the compression. That core is pinched far harder than the device's diagnostics assume, never relaxing to the pressure-balanced equilibrium from which its separatrix radius and temperature are conventionally read off. It also reproduces a non-axisymmetric, four-fold ($m=4$) deformation of the compressed core that appears in the experiment's end-on imaging but cannot form in the axisymmetric models used previously, a concrete sign that the formation phase carries genuinely three-dimensional, kinetic structure. The calculation is already practical on modest GPU hardware; as next-generation machines lift the remaining approximations and the fusion, ionization and radiation modules now in development come online, the same framework points toward end-to-end kinetic modeling of fusion-relevant devices.


\section*{Author Declarations}
\subsection*{Conflict of Interest}
The authors have no conflicts to disclose.

\section*{Data Availability}
The data that support the findings of this study are available from the corresponding author upon reasonable request.

\bibliographystyle{aipnum4-1}
\bibliography{refs}

\end{document}